\pdfoutput=1
\documentclass[aps, prl, twocolumn, amssymb, amsmath, superscriptaddress]{revtex4-1}
\usepackage{bm}
\usepackage{times}
\usepackage{graphicx}
\usepackage[colorlinks=true, letterpaper=true, pdfstartview=FitV, linkcolor=blue, citecolor=blue, urlcolor=blue]{hyperref}
\usepackage{array}
\usepackage{float}
\usepackage{colortbl}
\usepackage{multirow}
\usepackage{tabularx}

\begin{document}

\title{{\color{black}Hole-Doped Room-Temperature Superconductivity in H$_{3}$S$_{1-x}$Z$_x$ (Z=C, Si)}}
\author{Yanfeng Ge}
\affiliation{State Key Laboratory of Metastable Materials Science and Technology 
and Key Laboratory for Microstructural Materials Physics of Hebei Province,
School of Science, Yanshan University, Qinhuangdao, 066004, China}
\affiliation{Key Laboratory of Advanced Optoelectronic Quantum Architecture and Measurement, Ministry of Education, School of Physics, Beijing Institute of Technology, Beijing 100081, China}
\author{Fan Zhang}\email{zhang@utdallas.edu}
\affiliation{Department of Physics, University of Texas at Dallas, Richardson, Texas 75080, USA}
\author{Ranga P. Dias}
\affiliation{Department of Mechanical Engineering, University of Rochester, Rochester, New York 14627, USA}
\affiliation{Department of Physics and Astronomy, University of Rochester, Rochester, New York 14627, USA}
\author{Russell J. Hemley}
\affiliation{Department of Physics, University of Illinois at Chicago, Chicago, Illinois 60607, USA}
\affiliation{Department of Chemistry, University of Illinois at Chicago, Chicago, Illinois 60607, USA}
\author{Yugui Yao}\email{ygyao@bit.edu.cn}
\affiliation{Key Laboratory of Advanced Optoelectronic Quantum Architecture and Measurement, Ministry of Education, School of Physics, Beijing Institute of Technology, Beijing 100081, China}

\begin{abstract}
We examine the effects of the low-level substitution of S atoms by C and Si atoms 
on the superconductivity of H$_3$S with the $Im\bar{3}m$ structure at megabar pressures. 
The {\color{black}hole doping} can fine-tune the Fermi energy to reach the electronic density-of-states peak maximizing the electron-phonon coupling. 
This can boost the critical temperature from the original 203~K to 289~K and 283~K, respectively,  
for H$_3$S$_{0.962}$C$_{0.038}$ at 260~GPa and H$_3$S$_{0.960}$Si$_{0.040}$ at 230~GPa. 
The former may provide an explanation for the recent experimental observation of 
room-temperature superconductivity in a highly compressed C-S-H system [Nature \textbf{586}, 373-377 (2020)]. 
Our work opens a new avenue for substantially raising the critical temperatures of hydrogen-rich materials. 
\end{abstract}
\date{\today}
\maketitle

Pursuing room-temperature superconductors has been a major theme in condensed-matter and materials physics since the discovery of superconductivity in 1911. 
According to the Bardeen–Cooper–Schrieffer (BCS) theory, 
strong electron-phonon coupling and high phonon frequencies can conspire to produce superconductivity exhibiting high critical temperatures ($T_c$'s).
These two conditions can be achieved via strong covalent metallicity and low atomic mass, respectively. 
Naturally, metallic hydrogen and hydrogen-rich materials in general under pressure are plausible candidate high-$T_c$ superconductors~\cite{Ashcroft1968,Ashcroft2004,Tse2007,Chen2008,Eremets2008,Li2010,Gao2010,Zha2012,Wang2012,Feng2006}.
As a prime example, hydrogen sulfide H$_3$S with the $Im\bar{3}m$ structure has been theoretically predicted~\cite{Li2014,Duan2015} and 
experimentally confirmed~\cite{Drozdov2015,Einaga2016} to exhibit a maximum $T_c$ of $203$~K at $150$~GPa. 
{\color{black}Though a range of $T_c$'s has been reported in various studies~\cite{Drozdov2015,Einaga2016,Troyan2016,Huang2019,Shimizu2020,Mao2018}, 
the maximum $T_c$ of at least $180$~K for H$_3$S has now been reproduced in several different experiments~\cite{Drozdov2015,Einaga2016,Shimizu2020}.} 
This breakthrough has attracted a great deal of research interest~\cite{Bernstein2015,Errea2015,Papaconstantopoulos2015,Akashi2015,Nicol2015,Heil2015}. 
In particular, it has been predicted that under hole doping the $T_c$ of H$_3$S$_{0.927}$P$_{0.075}$ can be as high as $280$~K at $250$~GPa~\cite{Ge2016}. 
To date a variety of new hydrogen-rich materials with different structures have been proposed based on first-principles calculations~\cite{Peng2017,Liu2017,Tanaka2017,Ye2018,Cui2020,Sun2020}. 
As a paradigmatic system, the lanthanum superhydride LaH$_{10}$ with its novel hydrogen clathrate structure 
has been experimentally demonstrated to exhibit $T_c$'s of $250$--$260$~K at pressures of $170$--$185$~GPa~\cite{Somayazulu2019,Drozdov2019,Hong2020}.

A very recent experimental study has reported superconductivity in the C-S-H system at pressures of $140$--$275$~GPa, 
with the highest $T_c$ of $288$~K at $267$~GPa demonstrating room-temperature superconductivity~\cite{Snider2020}. 
Since the underlying crystal structure has yet to be determined,  pressure-induced structural changes over the measured range cannot be ruled out. 
However, the current experimental data are consistent with a continuous increase in $T_c$ with pressure 
but perhaps a discontinuous $\mathrm{d}{T_c}/\mathrm{d}P$ near $230$~GPa, 
suggesting a gradual structural deformation instead of an abrupt structural transition near that pressure.  
More experimental data are obviously required to establish the trend of $T_c$ versus pressure, and more importantly 
chemical analysis and X-ray diffraction are required to reveal the composition and crystal structure of the superconducting component in the C-S-H system.

Interestingly, the room-temperature superconductivity in the C-S-H system and its $T_c$ pressure dependence beginning below $200$ GPa~\cite{Snider2020} 
are strongly reminiscent of the predicted pressure effect on $T_c$ of hole-doped H$_3$S$_{1-x}$P$_{x}$ and H$_3$S$_{1-x}$Si$_{x}$~\cite{Ge2016}.
{\color{black}Here we systematically study the effects of C and Si substitution for S on the superconductivity of H$_3$S with the cubic $Im\bar{3}m$ structure at megabar pressures}  
based on first-principles calculations with the virtual crystal approximation (VCA)~\cite{Ge2016}. 
{\color{black} The cubic structure is examined here since it gives the highest $T_c$ for H$_3$S~\cite{Einaga2016}; the effects of substitution on phases derived from 
lower symmetry structures~\cite{Li2014,Goncharov2016,Akashi2016,Majumdar2017,Goncharov2017,Cui2020,Sun2020,Laniel} will be studied in future work.}
As demonstrated previously~\cite{Ge2016} and below, 
low-level substitution can fine-tune the Fermi energy to reach the electronic density-of-states (DOS) peak and thus maximize the electron-phonon coupling. 
It turns out that this can boost the $T_c$ from the original 203~K to 289~K and 283~K, respectively,  
for H$_3$S$_{0.962}$C$_{0.038}$ at 260~GPa and H$_3$S$_{0.960}$Si$_{0.040}$ at 230~GPa. 
The former may provide an explanation for the recent experimental observation of room-temperature superconductivity 
in the C-S-H system including its overall $T_c$ pressure dependence~\cite{Snider2020}. 
{The results thus also suggest an effect of C incorporation below $200$~GPa.}

\begin{figure*}[t!]
\centerline{\includegraphics[width=0.9\textwidth]{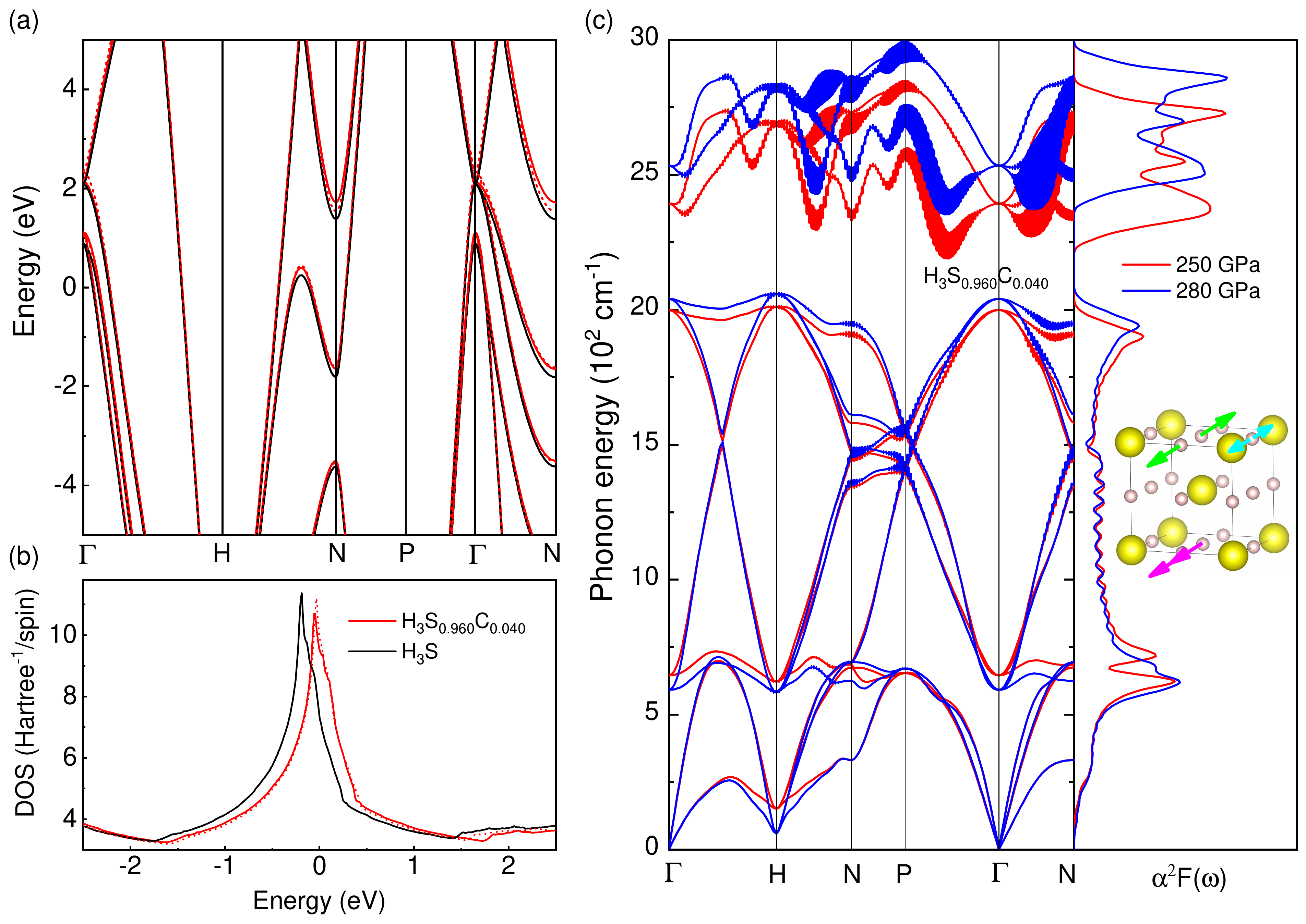}}
\caption{(a) Band structures and (b) DOS of H$_3$S at $220$~GPa (black), H$_3$S$_{0.960}$C$_{0.040}$ at $220$~GPa (solid red), 
and H$_3$S$_{0.960}$C$_{0.040}$ at $240$ GPa (dashed red), with their Fermi energies set to be zero. 
(c) Phonon dispersion and Eliashberg function $\alpha^2{F}(\omega)$ of 
H$_3$S$_{0.960}$C$_{0.040}$ at $250$ GPa (red) and 280 GPa (blue). The magnitudes of phonon linewidths are indicated by the line thickness. 
Inset: the low-frequency (magenta) and high-frequency (green) H-S bond-bending modes around $600$ $(2000)$~cm$^{-1}$ 
and the H-S bond-stretching modes around $2500$~cm$^{-1}$ (cyan).}
\label{fig1}
\end{figure*}

Following the approach introduced in Ref.~\onlinecite{Ge2016},
our calculations were performed within the framework of ABINIT~\cite{Gonze19971,Gonze19972,Gonze2005,Gonze2009} 
using the local-density approximation. 
The ion and electron interactions were treated with Hartwigsen-Goedecker-Hutter pseudopotentials~\cite{Hartwigsen1998}. 
The electronic ground-state properties were calculated on a $32\times32\times32$ Monkhorst-Pack $k$-mesh using the kinetic energy cutoff of $800$~eV.  
The phonon dispersions and the electron-phonon couplings were calculated on an $8\times8\times8$ $q$-grid 
using the density functional perturbation theory~\cite{Baroni2001}. The atomic substitution was simulated by the self-consistent VCA, 
where the virtual pseudopotentials of S$_{1-x}$Z$_x$ were set to be $V_{\rm VCA} = (1-x)V_{\rm S}+xV_{\rm Z}$.

The superconductivity of H$_3$S can be accurately described by the BCS theory, which underlines the aforementioned 
strong electron-phonon coupling and high phonon frequencies as the two most important factors in producing the high $T_c$ 
in this class of materials.  
Here we focus on the former factor, or equivalently the effect of the electronic DOS, for reasons that becomes clear below. 
{\color{black}H$_3$S has a DOS that reaches $7.43$~Hartree$^{-1}$/spin at its Fermi level~\cite{Papaconstantopoulos2015},} 
because of the presence of a van Hove singularity in the vicinity, 
as shown in Fig.~\ref{fig1}. By substituting the S atoms with C and Si, 
the H$_3$S system can be hole doped, and the Fermi level can be moved closer to the DOS peak,  
and by increasing the pressure, the DOS peak can be further enhanced. Both effects are illustrated in Fig.~\ref{fig1}(b). 
The dynamical stability of the crystal structure limits the substitution level to $x\leq 0.050$ for H$_3$S$_{1-x}$Z$_x$ (Z=C, Si) hereafter. 

Figure~\ref{fig1}(c) compares the phonon dispersions of H$_3$S$_{0.960}$C$_{0.040}$ at $250$ and $280$~GPa. 
The coupling of Fermi-level electrons with specific phonons is indicated by the thickness of the dispersion curves, 
i.e., the magnitudes of phonon linewidths 
\begin{equation}
\gamma_{\bm q\nu}=2 \pi \omega_{\bm q\nu}{\sum_{\substack{ij\bm{k}}}}{\vert{M^{\nu}_{i\bm k, j\bm{k+q}}}\vert}^2
\delta(\epsilon_{i\bm k}-\epsilon_{F})\delta(\epsilon_{j\bm{k+q}}-\epsilon_{F})\,,
\end{equation}
where $M^{\nu}_{i\bm k, j\bm{k+q}}$ are the microscopic electron-phonon matrix elements. 
High-frequency H-S (or H-Z) bond-stretching modes near $2500$ cm$^{-1}$ have the largest phonon linewidths, 
indicating strong electron-phonon coupling of the H-based vibrations. 
One such high-frequency phonon mode and two lower-frequency modes are illustrated in the inset of Fig.~\ref{fig1}(c).

\begin{figure*}[t!]
\centerline{\includegraphics[width=0.9\textwidth]{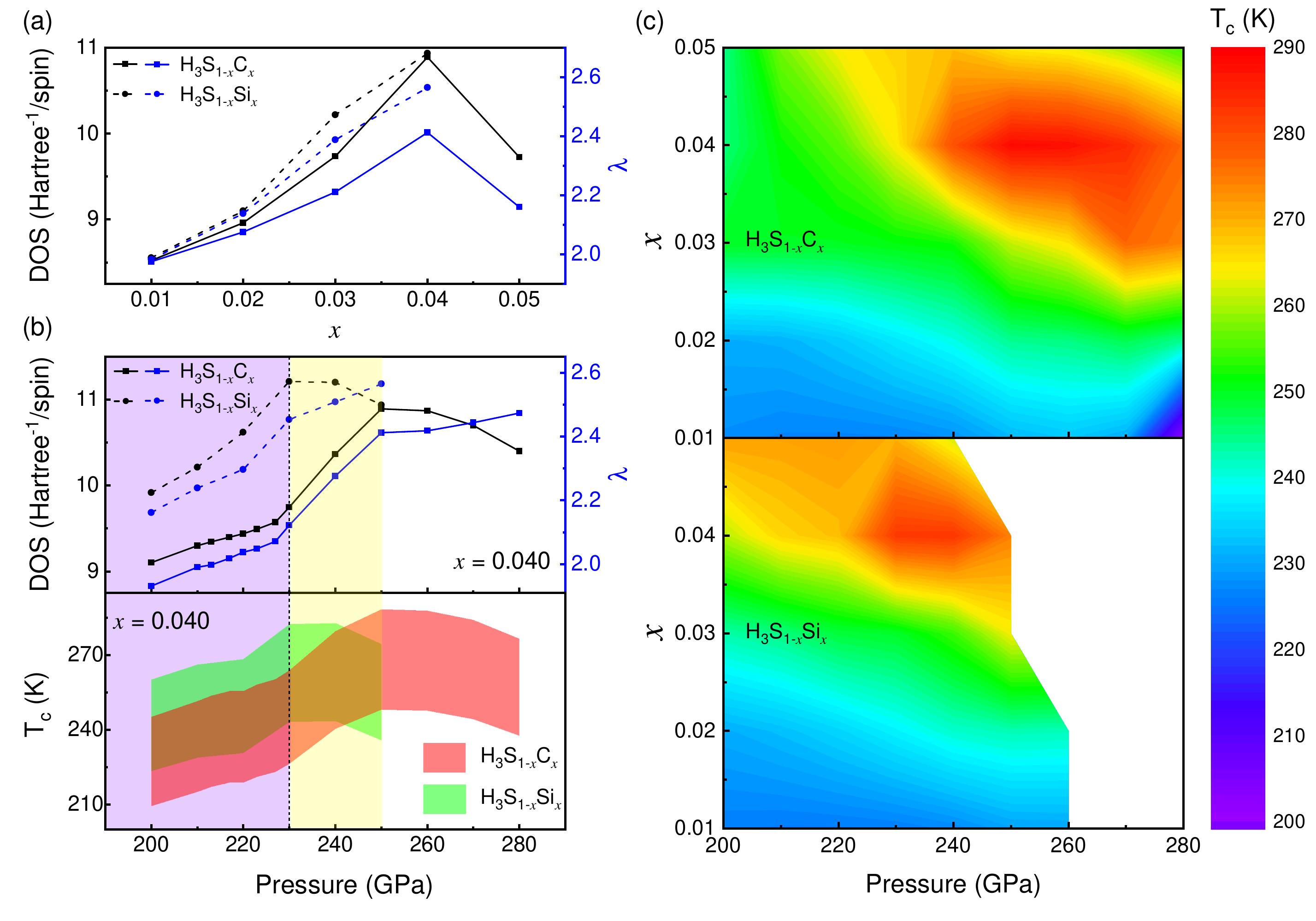}}
\caption{(a) DOS (black) and electron-phonon coupling $\lambda$ (blue) of cubic H$_3$S$_{1-x}$Z$_x$ (Z=C, Si) versus $x$ at $250$~GPa. 
{\color{black}Solid and dashed lines denote the results for H$_3$S$_{1-x}$C$_x$ (square) and H$_3$S$_{1-x}$Si$_x$ (circle), respectively.}  
(b) DOS (black), $\lambda$ (blue), and $T_c$ (shaded) versus pressure at $x=0.040$. 
For each $T_c$ curve, the upper and lower bounds are obtained by choosing $\mu^\ast=0.10$ and $0.15$, respectively. 
(c) The $T_c$ maps of H$_3$S$_{1-x}$C$_x$ and H$_3$S$_{1-x}$Si$_x$ versus substitution level and pressure for $\mu^\ast=0.10$. 
{\color{black}At high pressures or high substitution levels, cubic H$_3$S$_{1-x}$Si$_x$ becomes unstable; thus no data are shown.}}
\label{fig2}
\end{figure*}

The superconductivity of H$_3$S$_{1-x}$Z$_{x}$ (Z=C, Si) can be estimated by Eliashberg theory~\cite{Allen1983}, 
which takes into account the renormalization of electron-electron repulsion by electron-phonon interactions. 
This celebrated theory has been successfully used in predictions of superconductivity in hydrogen-rich materials 
as well as ambient-pressure conventional superconductors.  
Figure~\ref{fig2}(a) shows that for both C- and Si-substitutions at $250$~GPa  
the DOS at the Fermi level increases with increasing the substitution level $x$, reaches a maximum around $x = 0.040$, 
and then starts to decrease. This trend is similar to that found for H$_3$S$_{1-x}$P$_x$~\cite{Ge2016}, 
and as anticipated C- or Si-substitution is almost twice as efficient as P-substitution for hole doping. 
For both cases, the electron-phonon coupling $\lambda$ follows the trend of the DOS,  
as the influence of low substitution on the phonon frequency is weak and secondary. 
{\color{black}The upper-pressure limit of dynamical stability of cubic $Im\bar{3}m$ H$_3$S$_{1-x}$Z$_{x}$ varies with both Z and $x$. 
For example, the upper-pressure limit of H$_3$S$_{0.960}$C$_{0.040}$ is 280 GPa, and that of H$_3$S$_{0.960}$Si$_{0.040}$ is 250 GPa. 
Beyond these limits, acoustic phonons at the H point become imaginary signaling structural instabilities. 
As seen in Fig.~\ref{fig2}(c), our study focuses on pressures of $200$--$280$~GPa and the substitution levels up to $x=0.050$.  
At these pressures H$_3$S$_{1-x}$C$_{x}$ is found to be dynamically stable,  
whereas H$_3$S$_{1-x}$Si$_{x}$ becomes unstable beyond about $250$ GPa with a moderate $x$-dependence.}

As shown in Fig.~\ref{fig2}(b), for both C- and Si-substitutions at $x=0.040$, 
the DOS at the Fermi level increases first and then decreases as the pressure increases. 
The maximum DOS occurs at different pressures for the two cases, i.e., $250$~GPa for H$_3$S$_{0.960}$C$_{0.040}$ and $230$~GPa for H$_3$S$_{0.960}$Si$_{0.040}$. 
Notably, for C-substitution the DOS increases more markedly with pressure in the $230$--$250$~GPa range than that at lower pressures. 
A possible reason is the complex influence of compression on the electronic structure around the Fermi level, 
as implied in Fig.~\ref{fig1}(b) by the different DOS behavior at different pressures for a fixed doping level.
By contrast, the $\lambda$'s in both cases increase monotonically with pressure.  
The different DOS and $\lambda$ trends with pressure are likely due to softening of the phonons around $600$~cm$^{-1}$. 
As marked by the magenta arrows in the inset of Fig.~\ref{fig1}(c), 
these softened phonon modes, which also contribute to $\lambda$, are low-frequency H-S bond-bending modes~\cite{Errea2015}.  
As we see from $\lambda = 2\int\omega^{-1}\alpha^2\rm{F}(\omega)\rm{d}\omega$, the phonon softening is beneficial for enhancing $\lambda$. 
Indeed, there is a drop in the logarithmically averaged phonon frequency ${\langle\omega\rangle}_{\log}$ from $1364$~K at $250$~GPa to $1269$~K at $280$~GPa. 
This explains why $\lambda$ exhibits a modest upward trend while the DOS decreases with increasing pressure. 

Given the averaged phonon frequency ${\langle\omega\rangle}_{\log}$, effective Coulomb repulsion $\mu^\ast$, and electron-phonon coupling $\lambda$, 
the Allen-Dynes-modified McMillan formula~\cite{Allen1975,Durajski} 
\begin{eqnarray}
	T_c&=f_1f_2&\frac{\langle\omega\rangle_{\log}}{1.20}
	\exp{\left[-\frac{1.04(1+\lambda)}{\lambda-\mu^\ast(1+0.62\lambda)}\right]}
	\label{eq:tc}
\end{eqnarray}
can be implemented to predict the $T_c$. Here $f_1$ and $f_2$ are the strong coupling and shape correction factors~\cite{Allen1975}, respectively; 
a reasonable range of $\mu^\ast$ is between $0.10$ and $0.15$~\cite{range}. 
For H$_3$S at $200$~GPa~\cite{Drozdov2015}, our calculations yield the $T_c$ of $194$~K for $\mu^\ast=0.12$ and $203$~K for $\mu^\ast=0.11$. 
Following the behavior of the DOS, the $T_c$ increases faster in the $230$--$250$~GPa range than that at lower pressures, 
as shown in Fig.~\ref{fig2}(b). 
This behavior appears to parellel the upturn in $T_c$ above $230$~GPa observed experimentally in the C-S-H system~\cite{Snider2020}.
While the DOS drops with increase in pressure as discussed above, 
the phonon softening enhances $\lambda$ but weakens ${\langle\omega\rangle}_{\log}$. 
This implies a decrease of Debye temperature in the BCS theory. 
This suggests a maximum $T_c$ versus pressure that tracks the behavior of the DOS. 
In order to display the joint influence of hole doping and high pressure, as well as to identify the maximum $T_c$, 
we plot the map of the $T_c$ versus the substitution level and pressure for $\mu^\ast=0.10$ in Fig.~\ref{fig2}(c). 
In particular, the highest $T_c$ are $289$~K for H$_3$S$_{0.962}$C$_{0.038}$ at $260$~GPa and 
$283$~K for H$_3$S$_{0.960}$Si$_{0.040}$ at $230$~GPa. 
The former is very close to the highest $T_c$ of $288$~K at $267$~GPa observed in the C-S-H experiment~\cite{Snider2020}. 
{\color{black}Evidently in Fig.~\ref{fig2}(c), the highest $T_c$'s are reached near the structural instabilities. This appears to be consistent with the picture that 
soft phonon modes can be important for $T_c$ enhancement near structural instabilities 
in superconducting hydrides~\cite{Cohen1972,Quan2019,Chen2020}.}

In conclusion, we have examined the effects of hole doping on the superconductivity of H$_3$S 
with the $Im\bar{3}m$ structure at megabar pressures by using the first-principles calculations with the VCA. 
This fine-tunes the Fermi energy to reach the peak in the electronic DOS, maximizes the electron-phonon coupling, 
and boosts the $T_c$ to 289~K and 283~K, respectively, for H$_3$S$_{0.962}$C$_{0.038}$ at 260~GPa and H$_3$S$_{0.960}$Si$_{0.040}$ at 230~GPa. 
Because of the fewer valence electrons and the lighter atomic masses, 
the C- and Si-substitutions are more efficient in raising $T_c$ than substitution by P~\cite{Ge2016}. 
Although less stable at the higher pressure, Si-substitution raises the $T_c$ more than substitution by C below $240$~GPa. 
Most importantly, the C-substitution may provide an explanation for the recent experimental observation of room-temperature superconductivity 
in the C-S-H system and its $T_c$ pressure dependence above $200$~GPa~\cite{Snider2020}. 
Our findings indicate that hole doping in general--not limited to C-, Si-, and P-substitutions--is a robust approach to maximize the $T_c$ of H$_3$S. 
Looking forward, our study, together with Ref.~\onlinecite{Ge2016}, 
opens a new avenue for substantially raising the already high $T_c$'s of hydrogen-rich materials 
and calls for experimental investigation to systematically optimize the doping of these materials under pressure to reach still higher $T_c$'s. 

\begin{acknowledgments}
We are grateful to Roald Hoffmann and Eva Zurek for valuable comments on this work.
F.Z. is grateful to Anvar Zakhidov, Bing Lv, and Mikhail Eremets for valuable discussions at the initial stage of this work. 
The work is supported by the National Key R\&D Program of China (Grant No. 2020YFA0308800), 
the National Natural Science Foundation of China (Grants Nos. 11904312 and 11734003), 
the Strategic Priority Research Program of Chinese Academy of Sciences (Grant No. XDB30000000), 
the Project of Hebei Educational Department (Grants No. QN2018012), the UT Dallas Research Enhancement Fund, 
the US National Science Foundation (Grant Nos. DMR-1933622 and DMR-1809649), 
and the US Department of Energy (Grant Nos. DE-SC0020340 and DE-NA0003975).
\end{acknowledgments}

\end{document}